\documentclass[aps,prl,twocolumn,showpacs,groupedaddress,preprintnumbers]{revtex4}
\usepackage{amsmath, amsfonts,euscript,bbm}
\usepackage{ifthen}
\usepackage{graphicx}
\usepackage{psfrag}

\def\gsim{\mathrel {\vcenter {\baselineskip 0pt \kern 0pt \hbox{$>$} \kern 0pt \hbox{$\sim$} }}}

\def\pa{\partial}

\newcommand{\PRE}[1]{}
\newcommand{\Expect}[1]{\left\langle #1 \right\rangle}

\newcommand{\be}{\begin{equation}}
\newcommand{\ee}{\end{equation}}
\newcommand{\bea}{\begin{eqnarray}}
\newcommand{\eea}{\end{eqnarray}}

\newcommand{\nbox}{{\,\lower0.9pt\vbox{\hrule \hbox{\vrule height 0.2 cm
\hskip 0.2 cm \vrule height 0.2 cm}\hrule}\,}}

\def\lsim{\mathrel {\vcenter {\baselineskip 0pt \kern 0pt \hbox{$<$} \kern 0pt \hbox{$\sim$} }}}

\begin{document}

\preprint{UCI-TR-2008-16, MIFP-08-09, NSF-KITP-08-63, arXiv:0805.1229}

\title{
\PRE{\vspace*{1.5in}}
Tachyon Mediated Non-Gaussianity
\PRE{\vspace*{0.3in}}
}

\author{Bhaskar Dutta}
\email{dutta@physics.tamu.edu}
\affiliation{Department of Physics, Texas A\&M University, College Station, TX
77843, USA
\PRE{\vspace*{.5in}}
}

\author{Jason Kumar%
}
\email{kumarj@uci.edu}
\affiliation{Department of Physics and Astronomy, University of
California, Irvine, CA 92697, USA
\PRE{\vspace*{.5in}}
}
\author{Louis Leblond%
}
\email{lleblond@physics.tamu.edu}
\affiliation{Department of Physics, Texas A\&M University, College Station, TX
77843, USA
\PRE{\vspace*{.5in}}
}

\begin{abstract}
We describe a general scenario where primordial non-Gaussian
curvature perturbations are generated in models with extra scalar fields.
The extra scalars communicate to the inflaton sector mainly through the tachyonic (waterfall) field
condensing at the end of hybrid inflation.  These models can yield
significant non-Gaussianity of the local shape,
and both signs of the bispectrum can be obtained.
These models have cosmic strings and a nearly flat power spectrum,
which together have been recently shown to be a good fit to WMAP data.
We illustrate with a model of inflation inspired from
intersecting brane models.
\end{abstract}
\maketitle

\emph{Introduction.}
Single field slow-roll inflation generically predicts a near flat Gaussian
spectrum of fluctuations in the curvature of spacetime. Gaussianity is a consequence
of the slow-rolling conditions which require the inflaton to behave like a free field with negligible
self-interactions. We can parameterize the
non-Gaussianity (NG) in the curvature perturbation (denoted $\zeta$ here)
using the so called $f_{NL}$ formalism:
\begin{equation}\label{fnllocal}
\zeta(\vec{x},t) = \zeta_{Gauss} + \frac{3}{5}f_{NL} (\zeta_{Gauss}^2 - \Expect{\zeta_{Gauss}^2})\; ,
\end{equation}
where $\zeta_{Gauss}$ is the Gaussian piece of the curvature perturbation.
This ansatz for the curvature perturbations leads to the so-called local shape
for the bispectrum in Fourier space.

The recent five year analysis  from Wilkinson Microwave Anisotropy Probe
(WMAP5) \cite{Komatsu:2008hk} shows  $-9<f_{NL}<111$
while Yadav and Wandelt (using WMAP3 data)
have claimed a detection of primordial NG at
above 99.5\% confidence level \cite{Yadav:2007yy}.
It is expected that the upcoming Planck experiment should be able to detect $f_{NL}$ of order unity.
In the light of all these new and future measurements, it is very important to
investigate the origin of large non-Gaussianity in models of inflation.
It is also important to search for  predictions that are associated with
a larger $f_{NL}$ for other future observations, e.g., the shape of the
bispectrum, running of $f_{NL}$, etc.
The best  bound on $f_{NL}$ given above is for the local shape.
Another well known shape, the  equilateral shape, has also been constrained (using WMAP3 data)
to be $-256 < f_{NL}^{equil} < 332$
\cite{Creminelli:2006rz} while many other shapes
have not yet been searched for individually.

For single field smooth slow-roll models with canonical kinetic terms, $f_{NL}$ is of order of the slow-roll
parameters and is less than 1 \cite{Maldacena:2002vr}.
In this paper, we investigate a natural way to induce large non-Gaussianity from a ubiquitous feature
of string theory and models of physics beyond the standard model: extra scalars.
In the following, we will refer to the fluctuation
along the inflationary trajectory the
adiabatic mode while orthogonal fluctuations are entropy modes.
The entropy modes fluctuations are generically non-Gaussian
since in general their potential is not flat.
Given a mechanism to feed the entropy modes fluctuations into curvature perturbations,
the curvature spectrum will also be non-Gaussian.  In addition to this intrinsic NG of the entropy modes,
the non-linearity of the transfer mechanism between entropy modes and curvature perturbations will
also generate significant NG.

The scenario we present is general but can be easily realized
by intersecting brane models (IBMs) in string theory
\cite{Dutta:2007cr}. The IBMs are well-motivated since in these scenarios not only the standard model can
be constructed, but also the fermion mass hierarchies can be understood.  From the
effective field theory point of view, this scenario
gives a multi-fields generalization of D-term inflation \cite{Binetruy:1996xj}.
We will show that non-Gaussianity can be
generated at the end of inflation in this type of models from two main different sources. The first
source of NG is an ``intrinsic" component coming from self interactions of the entropy modes. The
second source is coming from the non-linear terms that arises when one transfers the entropy perturbations
into curvature perturbations.

The intrinsic component of non-Gaussianity in our work is similar to the
ref.~\cite{Bernardeau:2007xi} (building on previous work \cite{Bartolo:2001cw}) which also
considers  D-term inflation.
The non-linear local piece on the
other hand has been studied previously for multiple field hybrid inflation in \cite{Sasaki:2008uc} or in the context of modulated reheating \cite{Dvali:2003em}. 
But we find that in more general models, both contributions can appear and can be significant.
Moreover, we argue that loop-contributions to the bispectrum can also be important.
Thus, even simple, well-motivated models can yield a much more complicated picture
than previously discussed.

The intrinsic and non-linear non-Gaussian contributions both
have local shape with opposite signs that depend on the
hidden-sector charge of the extra
fields and on whether the dynamics of the
entropy mode is dominated by the gauge sector or the Yukawa sector of the theory.
Thus, we find in this setup that the  sign of $f_{NL}$ is not a stochastic
variable but rather fixed by the underlying physics of the system.
We find that the best regime to obtain large NG ($10\sim 100$) is the regime of D-term inflation
where we produce cosmic strings with tension $G\mu \sim 10^{-7}$ and where $n_s \sim 1$.
This is a good fit to the WMAP data as recently shown in \cite{Bevis:2007gh}.

\emph{The 2-pt Correlation Function.}
In this work, we will be interested in a model of hybrid inflation containing the usual
inflaton $\phi$ and the tachyonic (waterfall) field $T$ but extended with an extra scalar
field $\chi$.  During inflation, the tachyon has a positive mass and its VEV is pinned at
zero. Below some critical value ($\phi_c$) of the inflaton, the mass of the tachyon becomes
negative and it condenses, ending inflation.  We will assume that the $\chi$ field has
low mass, fluctuates during inflation but does not directly couple to the inflaton
field. Instead its only effect is to modulate the mass of the tachyon and to change the critical
value at which inflation ends.  In some ways, the tachyon acts as a messenger between the hidden
sector field $\chi$ and the inflaton field $\phi$. The action for this system is of the
following form
\bea
S & = &\frac12 \int\sqrt{g}[M_p^2 R - (\partial\phi)^2- (\partial T)^2 - (\partial\chi)^2 -2 V]\; , \nonumber\\
V &=& V_{\rm{inf}}(\phi) + V_{\rm{hid}}(\chi) + V_{\rm{mess}}(\phi,\chi,T)]\; .
\eea
During inflation, the tachyonic field has vanishing VEV and the messenger part of the
potential $V_{\rm{mess}}$ vanishes as well (all coupling in this part of the potential
involves the field $T$). Assuming that the hidden sector potential contribute a negligible amount
to the vaccuum energy, the Hubble scale during inflation is equal to
$H^2 = \frac{V_{\rm{inf}}(\phi)}{3M_p^2}$.  

This model is of the form considered in \cite{Bernardeau:2007xi, Sasaki:2008uc}.  Using the $\delta N$
formalism \cite{Sasaki:1995aw}, we can compute the curvature perturbation $\zeta(\vec{x},t)
= \delta N$ where $N$ is the number of efolds.
Since the inflationary potential is only a function of $\phi$, the field $\chi$ only affects
the critical value at which inflation ends.  The fluctuation in the number of
efolds when $\phi \rightarrow \phi + \delta\phi$ and $\chi \rightarrow \chi + \delta\chi$ is
\bea\label{deltaN}
\delta N &= & \left.\left.-\frac{H}{\dot\phi}\delta\phi\right|_* + \frac{H}{\dot\phi}
 \frac{\partial\phi_c}{\partial\chi} \delta\chi\right|_{\phi_c} \\
&& \left.+ \frac12\frac{H}{\dot\phi}\frac{\partial^2\phi_c}{\partial\chi^2}
\left(\delta\chi^2 -\Expect{\delta\chi^2}\right)
\right|_{\phi_c}
+ \cdots \; ,\nonumber
\eea
where the subscript $*$ means evaluated at horizon crossing. We have kept only the leading
term at second order and we substracted a number
to ensure that $\Expect{\delta N} = 0$.
We will find that non-Gaussianities are generated both
at linear order in the variable $\delta\chi$ as well as from the quadratic term. In curvaton models
\cite{Lyth:2001nq} (or more recently
in ekpyrotic models \cite{Buchbinder:2007at}),
NG is generated from these second order terms.  The validity of the $\delta N$ expansion to higher
orders and the NG that are generated from these terms were first discussed in \cite{Lyth:2004gb}. We define
\bea
\gamma &\equiv & \left. \frac{\pa \phi_c}{\pa \chi}\right|_{\phi_c}\; ,
\eea
and this function encodes the coupling between $\chi$ and $\phi$ through the tachyonic field.
To evaluate the 2-pt function we use the standard result that $\Expect{\delta\phi\delta\phi}\sim H^2/4\pi^2$.
Note that both $\delta\phi$ and $\delta\chi$ are not exactly constant outside of the horizon and they will
decay exponentially because of their mass terms with $ \left.\delta\chi\right|_f \sim \frac{H_*}{2\pi} \kappa$
where
\bea\label{damping}
\kappa  & = & e^{-\int_{t_*}^{t_f} \eta_\chi Hdt} \sim e^{-\eta_\chi N_e}\; ,
\eea
and we used the following definitions for the slow-roll parameters.
\begin{align}
&\epsilon =-\frac{\dot H}{H^2}, & \eta_\varphi = \frac{V_{,\varphi\varphi}}{3H^2}, && \eta_\chi
= \frac{V_{,\chi\chi}}{3H^2}\; .
\end{align}
Since we need to evaluate $\delta\chi$ at the end of inflation, significant non-Gaussianity
will only arise if this damping
is not too large, requiring $\eta_{\chi}$ to be small.  However, $V_{,\chi}$ need not be
particularly small, which distinguishes $\chi$ from the inflaton.
The 2-pt curvature correlation function is therefore \cite{Bernardeau:2007xi}
\bea\label{2pt}
\Expect{\zeta(\vec{k}_1) \zeta(\vec{k}_2)} &=& (2\pi)^3\delta^3
(\vec{k}_1+\vec{k}_2)2\pi^2\mathcal{P}_2^\zeta k^{-3}\; ,\nonumber\\
  \mathcal{P}_2^\zeta & = &
{H^2_* \over 8\pi^2 M_{pl} ^2} \left({1\over \epsilon_*}
+{\gamma^2 \kappa^2 \over \epsilon_f} \right)\; ,
\eea
where $f$ means
evaluated at the end of inflation. In the last expression, we have not included loop contributions
coming from the higher order terms in the $\delta N$. They can be made small as we will discuss in more
detail later on.
The spectral index is easily computed from $n_s -1 = \frac{d\ln \mathcal{P}_2^\zeta}{d\ln k}$, the
formula simplifies in the limit where $\epsilon_* \approx \epsilon_f$,
\bea
n_s -1 & = & \frac{1}{1+\gamma^2\kappa^2}\left(-6\epsilon_* + 2\eta_\varphi+ \gamma^2\kappa^2
(-2\epsilon_* + 2\eta_\chi )\right)\; .\nonumber
\eea

\emph{3-pt Correlation Function.}
The 3-pt correlation function can also be calculated using the same formalism although in this
case we can neglect the part coming from the inflaton $\phi$ since it is going to be suppressed
by power of the slow-roll parameters \cite{Maldacena:2002vr}.
The field $\chi$ on the other hand can be non-Gaussian for a general potential $V_{\rm{hid}}(\chi)$,
and this non-Gaussianity can be transferred to the curvature spectrum.  We will
denote by $\Expect{\zeta^3}_{\rm{int}}$ the piece of the bispectrum which is intrinsic
to the field $\delta \chi$ and arises directly from its higher-order self-interactions.  This
non-Gaussianity in $\delta \chi$ is transferred to the $\delta N$ spectrum by the linear
terms in eq. (\ref{deltaN}), yielding
\be\label{3ptcurvature}
\Expect{\zeta^3}_{\rm{int}} =\left.\left( \frac{H}{\dot\phi}\right)^3\gamma^3
\Expect{\delta \chi^3}\right|_{f}\; .
\ee
One can calculate this term using the in-in formalism \cite{Maldacena:2002vr} with
an interaction
Hamiltonian simply given by $H_I = \int d^3\vec{x} \sqrt{-g} V_{\rm{hid}}(\chi)$. This computation
was first done in \cite{falk} (see also \cite{ Bernardeau:2007xi, again} for more recent derivations).
The result is:
\bea\label{bispectrum}
&&\Expect{\zeta_{\vec{k}_1} \zeta_{\vec{k}_2}\zeta_{\vec{k}_3}}_{\rm{int}} =  (2\pi)^3
\delta^3(\sum \vec{k}_i)  \frac{H^2  V_{,\chi\chi\chi}\kappa^6} { 8 \sqrt{2} M_{pl}^3
\epsilon_{f}^{3\over 2}\prod k_i^3} I_3\; ,
\nonumber\\
&&I_3 = \frac49 k_t^3 - \frac13 \sum k_i^3(\gamma_e - N_e) - \sum_{i<j}k_ik_jk_t\; .
\eea
where $\gamma_e$ is the Euler number and we have defined $k_t = \sum k_i$.  The number of efolds is
$N_e = -\log(\frac{k_t}{a_f H_f})$ and we have used the fact that in the slow-roll 
regime $\frac{H}{\dot\phi} =
-\frac{1}{\sqrt{2\epsilon}M_p}$.
Since the number of efolds from horizon crossing is around 60 from observations,
we see that the dominant term is $I_3 \sim \frac{N_e}{3} \sum k_i^3$. This is exactly the local shape
obtained from Eq. (\ref{fnllocal}) (neglecting the weak momentum dependence in $N_e$).
Indeed, Fourier transforming $\zeta$ in (\ref{fnllocal}), one can compute the local 3-pt function to be
\be
\Expect{\zeta_k^3} = (2\pi)^7\delta^3(\sum\vec{k}_i) \frac{3}{10} f_{NL} (\mathcal{P}_2^\zeta)^2 \frac{\sum k_i^3}
{\prod k_i^3}\; .
\ee
Comparing the last two expressions we find the following intrinsic contribution to the $f_{NL}$.
\bea\label{fnleff}
f_{NL}^{\rm{int}} &= &
{5\over 9\sqrt{2}} {N_e M_p \gamma^3 \kappa^6 \over H^2 \epsilon_{f} ^{3/2}}
\left({1\over \epsilon_{*}} +{\gamma^2 \kappa ^2\over \epsilon_f} \right)^{-2}
V_{,\chi\chi\chi}\; .
\eea
where we use the subscript ``$\rm{int}$'' to remind the reader that this NG is coming from
self-interactions of the entropy modes. This kind of intrinsic NG occurs in models of inflation 
with non-trivial kinetic terms and they usually have very distinctive shapes in momentum space 
 (see for example \cite{Babich:2004gb, Chen:2006nt}). 
 In our case, the 3-pt function is dominated by the term proportional to the number of efolds which 
 originates from the classical evolution of the field outside of the horizon and it therefore has the local shape. We should note that this type of intrinsic NG (with local shape) also occurs in model of
non-local inflation where similar self-interactions are important \cite{Barnaby:2006hi}.

In addition to the intrinsic NG, the non linear terms in Eq.~(\ref{deltaN}) will
also contribute to the bispectrum.  In this case, Gaussian fluctuations in
$\delta \chi$ are transferred non-linearly to
$\delta N$, inducing non-Gaussianity.  Because the non-linear terms in the $\delta N$ expansion
are local in space-time, they will give a contribution of the
local form and the leading term is
\be\label{loc}
f_{NL}^{\rm{loc}} = - \frac{\partial \gamma}{\partial \chi} \frac{5 \gamma^2\kappa^4 M_p}{3\sqrt{2}
\epsilon_f^{3/2}} \left({1\over \epsilon_{*}} +{\gamma^2\kappa^2 \over \epsilon_f} \right)^{-2} + \cdots
\ee
where we have neglected higher order terms in the $\delta N$ expansion. Note that these two contributions
have the same shape but opposite signs.
The ratio of the intrinsic contribution to $f_{NL}$ compared to the local contribution is
\be
\beta \equiv  \left|\frac{f_{NL}^{\rm{int}}}{f_{NL}^{\rm{loc}}}\right| =
\frac{1}{3}\frac{\gamma}{\gamma_{,\chi}}\frac{V_{,\chi\chi\chi}}{H^2} N_e \kappa^2
\ee
For the case where $\gamma$ is a constant (as in
\cite{Bernardeau:2007xi}), $f_{NL}^{\rm{loc}}$ vanishes and the intrinsic
contribution is the most important one.  For the case where $\gamma$ is a linear
function of $\chi$, we have $\gamma_{,\chi} \sim \gamma/\chi$ and the ratio can be simplified to
\be
\beta \sim  \eta_\chi N_e \kappa^2
\ee
up to an order 1 coefficient which depends on the exact form of the potential. Given that
one must have $\eta_{\chi} < N_e^{-1}$ in order for the damping (c.~f.~ Eq.~(\ref{damping})) to be
small, this ratio is usually smaller than 1 although for values of $\eta_\chi \sim 0.01$, it can
be of order 1.

Finally there are higher orders (loop) contributions to the 2-pt and 3-pt
that we have not included here. For example, from Eq (\ref{deltaN}), there will
be a correction to the 2-pt curvature function of the following form
\be\label{3ptcorrection}
\Expect{\zeta^2_k}_{\rm{loop}} \supset \left(\frac{H}{\dot\phi}\right)^2 \gamma_{,\chi}\gamma \Expect{\delta\chi_k \int
\frac{d^3k'}{(2\pi)^3} \delta\chi_{k - k'}\delta\chi_{k'}}\; .
\ee
This is a loop factor and the integral is IR divergent with a log that must be regulated. A natural
cutoff is to use a box of the size of the universe today \cite{Lyth:2007jh}. Then the loop factor is
of order $\frac{\ln(kL)}{2\pi^2} \sim \frac14$ and using the intrinsic 3-pt function of $\delta\chi$
computed previously, we get a term of the order of
\be\label{4ptcorrection}
 - {H^2_* \over 8\pi^2\epsilon_f M_{pl} ^2} \gamma_{,\chi}\gamma N_e \kappa^6
V_{,\chi\chi\chi} \frac{\ln(kL)}{2\pi^2}\; .
\ee
For a given model, it is possible for this term to dominate over the $\gamma^2$ term in Eq. (\ref{2pt})
although for the perturbation theory to remain under control, one should require that $\gamma$ is small
enough to insure that $\delta\phi^2$ is the dominant term in the 2-pt.  In addition one can get an additional contribution to the 2-pt proportional to
\be
 \left(\frac{H}{\dot\phi}\right)^2 \gamma_{,\chi}^2 \int
\frac{d^3k' d^3k''}{(2\pi)^6} \Expect{\delta\chi_{k - k'}\delta\chi_{k'} \delta\chi_{k - k''}\delta\chi_{k''}}\; .
\ee
This would give a contribution proportional to the square of the 2-pt function
(of the order of $ \frac{H^4}{\epsilon_f M_p^2} \gamma_{,\chi}^2\kappa^4\ln kL$) as well as
a possible 2-loop contribution proportional to the 4-pt function. If we assume that
$\gamma_{,\chi\chi}$ is very small, we can stop here as all subsequent terms would be negligible.
A systematic approach to calculate all these loops contributions has been developed
recently in \cite{Crocce:2005xy}. Similar terms contribute to the 3-pt
function as well \cite{Zaballa:2006pv}. A more thorough analysis of these
additional contributions is important and we leave a detailed calculation of these
terms for further work. For the example we discuss later, we parametrically estimated all the loop contributions and chose a point in the parameter space where we can neglect the loops. 
Nevertheless, we need to point out that we also found region in the parameter space
where these loops corrections must be included and can even dominate.

Some comments are in order about the validity of the perturbative analysis. Assume for the moment
that $\gamma_{,\chi} \approx 0$ such that only $f_{NL}^{\rm{int}}$ is non-zero.
Then, in order to be able to compute the 2-pt correlation function of $\delta\chi$ using only
the quadratic action like we did, the interaction terms must be small \cite{Leblond:2008gg}.
In our context, it is enough to demand that $V_{,\chi\chi\chi} < H$,
or equivalently,
\bea\label{perturbative}
\gamma &>&
2 (P_2^\zeta)^{1/6} \left(\frac{f_{NL}^{\rm{int}}}{N_e}\right)^{1/3}\; ,
\eea
where
we used Eq. (\ref{fnleff}) and we assumed $\epsilon_*\sim \epsilon_f$ and $\gamma < 1$.
This tells us that in order to obtain a large $f_{NL}$ we cannot
take the coupling between the two sectors ($\chi$ and $\phi$) to be arbitrarily weak.  Indeed,
the smaller $\gamma$ corresponds to a more strongly interacting scenario for a given $f_{NL}$.
Note that this is a computational limit, it might be very interesting to have
a strongly coupled hidden sector. We leave this to future investigation.  In the case 
where $\gamma_{,\chi}$ is non zero then one must ensures that the loop factors discussed above 
are small corrections to the 2-pt function. This can be achieved by taking $\gamma$ sufficiently small.

\emph{A multi-field scenario.}
We illustrate this method of generating non-Gaussian curvature
perturbations with a slightly modified version of a $D$-term
inflationary scenario arising in IBMs \cite{Dutta:2007cr}.
This IBM scenario uses the inherent
extra gauge symmetry of the hidden sector to reduce the amount of
fine-tuning required for inflation, and can accommodate Standard Model embeddings
with interesting phenomenology.
We will find that this well-motivated scenario will exhibit 
the rich structure of non-Gaussian perturbations which we have discussed 
above.

The key feature of this inflationary
setup is that we have 4 different D-brane stacks with
$U(1)_{\rm{inf},1,2,3}$ gauge
theories on their world volumes.
There are several chiral multiplets which live at the
topological intersections of any two branes with each other and
transform in bifundamental of the gauge groups on the two branes.
The resulting D-term potential and superpotential are
\bea\label{dterm}
V_{\rm{inf}} ^D &=& {g^2 \over 2} (|\phi_{+}| ^2
-|\phi_-| ^2 -|\phi_{NG}|^2+ \cdots-\xi)^2\; ,
\nonumber\\
V_{\rm{rest}} ^D &=&
{g_1 ^2 \over 2} (|\phi_-| ^2 -|S| ^2 + \cdots -\xi_1 )^2
\nonumber\\
&&+{g_2 ^2 \over 2} (|S| ^2 -|\phi_{+} | ^2 +q|S_{NG}|^2 + \cdots-\xi_2 )^2
\nonumber\\
&&+{g_3 ^2 \over 2} (|\phi_{NG}|^2 -q|S_{NG}|^2)^2 +\cdots.\; ,\nonumber\\
W &=&\lambda S \phi_{+}   \phi_{-}
+ \lambda_{NG}  S_{NG} \phi_{+}  \phi_{NG}\; ,
\eea
where $\cdots$ denotes additional fields which arise at the intersections
of these 4 D-brane stacks with other branes which are not relevant for us.
$V_{\rm{inf}} ^D$ is the D-term potential of the brane which generates inflation,
and $q=\pm 1 $ is the charge of $S_{NG}$ under $U(1)_2$ (in an IBM construction,
this sign is determined by the orientation of topological intersection at which the
multiplet lives).  Note that we have
made a particular choice of the sign of the charge for all other fields, but
we have left the charge of $S_{NG}$ undetermined because we will find that
it has interesting observational consequences.
The superpotential will
generate F-terms (setting $\phi_- = \phi_{NG}=0$)
\be
V^F = \lambda^2 S^2\phi_+^2 + \lambda_{NG}^2 S^2_{NG}\phi_+^2\; .
\ee
Without any loss of generality, we assume $\xi >0$.  As a result,
$\phi_{+}$ is
the waterfall field.  $S$ is the inflaton; its vev gives
$\phi_+$ positive mass. $S_{NG}$ is the new field added compared to the set-up in \cite{Dutta:2007cr}
and this will be our field $\chi$.  On the D-flat
direction, $V^{D}_{\rm{rest}}=0$ and $V_{\rm{inf}}^D =\frac{g^2}{2} \xi^2$.
An inflaton potential is induced from the one-loop Coleman-Weinberg potential
\bea
V_{\rm{inf}} & = & {g^2 \xi^2 \over 2}
\left[1+{g^2 \over 16\pi^2}V_{CW}(x)
\right]\; ,\nonumber\\
V_{CW}(x) & = & (x^2+1)^2\ln(x^2+1)-2x^4\ln x^2\nonumber\\
&& + (x^2-1)^2\ln(x^2-1) -4\ln2\; ,
\eea
where $x = \frac{\lambda^2 \phi^2}{g^2\xi}$ and we
have neglected $\chi$ contributions to the 1-loop potential
(in the regime we will consider, they are small).
Using $U(1)$ rotational invariance, we may define
\begin{align}
&\phi \equiv  \rm{Re} \,S, &
\chi \equiv  \rm{Re} \,S_{NG}, &&
T \equiv \rm{Re} \,{\phi_+}\; ,
\end{align}
and we can set all the imaginary parts to zero.
The hidden sector field $\chi$
has a quartic coupling given by
$V_{\rm{hid}} = {\nu^2 \over 4} \chi^4= {g_2 ^2 + g_3 ^2 \over 2} \chi^4$.
Finally the mass of the tachyonic field depend on the quadratic sum of two fields
\be
m_{T} ^2 = -g^2 \xi + \lambda^2 \phi^2
+(\lambda_{NG}^2-q g_2 ^2) \chi^2\; .
\ee
Inflation ends when the inflaton reaches the critical value $\phi_c^2 =  \frac{g^2 \xi}{\lambda^2}
+ \frac{(qg_2 ^2- \lambda_{NG}^2)\chi^2}{\lambda^2}$ which gives
\bea
\gamma &=&\left.\frac{\pa \phi_c}{\pa \chi}\right|_{\phi_c}
= {qg_2 ^2 - \lambda_{NG}^2 \over \lambda^2}
{\overline{\chi} \over \phi_c}\; .
\eea
We denote  the stochastic part of the field $\chi$ by $\overline{\chi}$,
which is a random variable with a mean of zero.
The quantum perturbations of $\chi$ grows to be of order $H$ at horizon exit.
At this point, one can assume a semi-classical treatment where the field $\chi$ is
undergoing a random walk with steps of order $H$ for each interval of time $1/H$.
The Fokker-Planck equation then gives
the non-Gaussian probability distribution for $\overline{\chi}$.
We  thus find $\overline{\chi} = \sigma H/ \sqrt{\nu}$, where
a typical range is $-0.6 < \sigma  < 0.6 $ \cite{Bernardeau:2007xi}.
Note that $\gamma$ depends on this stochastic parameter and this is one of the main difference
from \cite{Bernardeau:2007xi}.

\emph{Analysis.}
At this point the system has been reduced to D-term inflation.
In our analysis we  will take the limit
${\xi \over \lambda^2} \gg \frac{N_e M_p^2}{2\pi^2}$ and assume that $\gamma \ll 1$ as well as
$\phi_c^2 \sim \frac{g^2 \xi}{\lambda^2}$.
In this limit,
$\phi_{*} ^2 \sim \phi_{f} ^2 \sim {g^2 \xi  \over \lambda^2}$ (or equivalently
$\epsilon_* \sim \epsilon_f$),
and the inflaton does not move much during inflation.
We must go to this limit in order to realize significant non-Gaussianity,
because the perturbations to $\delta N$ can arise both from
fluctuations at the start of inflation and from the end.
The initial fluctuations (in our model) are
Gaussian, so significant non-Gaussianity can only arise if the
non-Gaussian perturbations at the end of inflation are of about the
same scale.  This also means that the spectral index will be close to
one. In this regime,the 2-pt. normalization is
\be
P_\zeta \sim 2.28 \times 10^{-9} \sim \frac{4\pi^2}{6}\frac{\xi^3}{(2\ln(2))^2\lambda^2M_p^6}\; .
\ee
The local and intrinsic contributions to $f_{NL}$ are
\bea
f_{NL}^{\rm{int}} &\sim& \frac{5\ln(2)}{2\pi^2}\frac{\nu^2 N_e \kappa^6\overline{\chi}^4
(qg_2 ^2 -\lambda_{NG}^2 )^3M_p^4}
{\lambda^2\xi^4 g^4}\; \nonumber\\
f_{NL}^{\rm{loc}} &\sim&  -\frac{ 5 \ln(2)}{12\pi^2}\frac{\overline{\chi}^2
(qg_2 ^2 -\lambda_{NG}^2 )^3\kappa^4M_p^2}
{\lambda^2\xi^2}
\eea
The sign of $f_{NL}$ is determined by the sign of
$qg_2 ^2 -\lambda_{NG}^2 $.
Note that this dependence arises from the
way these couplings affect the mass of the waterfall field;
the Yukawa coupling adds a positive mass contribution and
delays the end of inflation, while the gauge coupling term
adds a mass contribution whose sign depends on the charge
$q$, and can either delay or hasten the end of inflation.
In particular, if $q=+1$ and $g_2^2 > \lambda_{NG}^2$, then
$qg_2 ^2 -\lambda_{NG}^2 >0$.  If this model arises from the
type of IBM construction described in \cite{Dutta:2007cr}, then one
would indeed expect $g_2^2 > \lambda_{NG}^2$, because Yukawa
couplings are exponentially
suppressed (in Type IIA string theory, they arise from worldsheet 
instatons).  On the other hand, if $q=-1$, then we can
set $\lambda_{NG}=0$, because the Yukawa term
$\lambda_{NG} S_{NG} \phi_+ \phi_{NG}$ is not classically
gauge-invariant (both $\phi_+$ and $S_{NG}$ would have charge
$-1$ under $U(1)_2$).  We then find that $qg_2 ^2<0$.  We thus
directly see how the sign of the charge of the new scalar determines
the sign of the various contributions to $f_{NL}$ in this model.

There are a few  consistency conditions one must satisfy;
we take $\eta_{\chi } \lsim 0.01$ so that the
non-Gaussian perturbations in $\chi$ do not decay significantly
between horizon-crossing and the end of inflation.
For perturbative control we must satisfy Eq. (\ref{perturbative}).
As a toy example, we take $\xi   \sim  2.9 \times 10^{-6}M_{pl}^2$,
$\lambda \sim  2\times 10^{-4}$, $qg_2 ^2 -\lambda_{NG}^2  \sim - 0.17$,
$\nu \sim 0.05$, $g^2 \sim  10^{-3}$ and
$\overline{\chi} \sim 0.12H / \sqrt{\nu}$.
We then satisfy the above consistency conditions
and generate $\gamma \sim - 0.5$ and
\begin{align}
&f_{NL}^{\rm{int}} \sim -8 \; ,
&n_s \sim 1.002\; , \nonumber\\
&f_{NL}^{\rm{loc}} \sim 45\; ,
&G\mu \sim  7 \times 10^{-7}\; .
\end{align}
where we included the cosmic strings produced in such models (with $G\mu = \xi/4$).
For this point in parameter-space, one can check that the
loop corrections Eq. (\ref{3ptcorrection}, \ref{4ptcorrection})
scales as $N_e \nu^2$ and $H^2/\chi^2$ respectively and for the parameters chosen above, they are small.
But clearly for other points in parameter space one or both of these expansions
parameters could be significant, in which case one would have to compute higher-order
terms as well.

As is typical in D-term inflation, this model predicts cosmic strings.
When their contribution to
the power spectrum is correctly accounted for, $n_s \sim 1$ can be completely consistent with
WMAP data \cite{Bevis:2007gh}.
One expects that larger values of $\xi$ can lead to smaller values of
$n_s$ (though if the resulting cosmic string tension exceeds observational bounds, the
strings must be unstable).  In this case, large values of $f_{NL}$ would require
larger values of $qg_2 ^2 -\lambda_{NG}^2$.
We can also find points in the parameter space with large $f_{NL}$
where $G\mu$ is smaller (and satisfy the bound of \cite{Pogosian:2003mz}) but in this case,
the WMAP bound on $n_s$ while not excluding our model, does not favor it. Also, we find that
the smaller the cosmic string tension the harder it becomes
to satisfy all the consistency conditions and yet get large $f_{NL}$.
We have assumed that the vev of the inflaton is sub-planckian.  This
implies that the energy density in gravity waves will be unobservably
small.  Significant gravity waves can be generated if the inflaton
traverses super-Planckian scales, but in that limit the effective
field theory is no longer valid. Finally, we should mention 
that a different mechanism for generating 
NG from the tachyon was presented in \cite{Enqvist:2005qu}.
 
\emph{Discussion.}  We have illustrated a simple model for generating
primordial non-Gaussian curvature perturbations.  This model can arise
easily from intersecting brane models and
allows a large $f_{NL}$ with either sign which
one expects to settle soon from the results of WMAP and Planck.
The shape of the non-Gaussianity is local. One can fit the CMB data with
large non-Gaussianity and a flat spectrum
once the contribution from cosmic strings is included correctly. Without the cosmic strings, the
current bound on $n_s$ implies this model is
not favored by the data but not excluded. The fact that the tachyon mass depends on additional fields is a
feature shared by
many string theory models of inflation \cite{Leblond:2006cc}.

In this general and well-motivated effective field theory model,
we have found a relatively complicated structure to the non-Gaussian
contributions, with contributions from both intrinsic and non-linear
terms, and possible loop-contributions as well.  The non-linear
contributions are usually dominant.
We have found that the sign of $f_{NL}$ depends largely on
the hidden-sector charge
of the extra field, and on the relative importance of the gauge and
Yukawa sectors.

Another model of D-term inflation in
string theory was
recently reanalysed
in \cite{Haack:2008yb} (and effects from entropy modes were discussed in \cite{Brandenberger:2008if}) and
it will be interesting to study whether NG can also be generated in their model. An important difference
in the D3/D7 model compared to the IBM model we presented is the fact that in the former, the Yukawa coupling
and gauge coupling are the same. Therefore to achieve the $n_s \sim 1$ regime, they need a very small gauge
coupling (of order $10^{-4}$) while in our case it can be larger.  Of course, the D3/D7 model contains a full
string theory description with backreaction and moduli stabilization effects included.  It would be interesting
to see if NG of the type presented in this work still survives in a more concrete model from string theory.

We are grateful to R. Brandenberger, K. Dasgupta, B Garbrecht, M. Kaplinghat, E. Komatsu,
 R. Holman, A. Linde, S. Shandera and M. Wyman  for useful discussions,
and to the KITP for its hospitality. We are particularly thankful to D. Lyth for pointing out to us the
importance of the local contribution to the NG.  This work is supported in part by DOE Grant
DE-FG02-95ER40917, NSF Grants No.~PHY--0239817, PHY--0653656, PHY05-51164 and PHY--0505757.


\begin{thebibliography}{10}

\bibitem{Komatsu:2008hk}
  E.~Komatsu {\it et al.}  [WMAP Collaboration],
  arXiv:0803.0547 [astro-ph].

\bibitem{Yadav:2007yy}
  A.~P.~S.~Yadav and B.~D.~Wandelt,
  arXiv:0712.1148 [astro-ph].

\bibitem{Creminelli:2006rz}
  P.~Creminelli, L.~Senatore, M.~Zaldarriaga and M.~Tegmark,
  JCAP {\bf 0703}, 005 (2007)
  [arXiv:astro-ph/0610600].

\bibitem{Maldacena:2002vr}
  J.~M.~Maldacena,
  JHEP {\bf 0305}, 013 (2003)
  [arXiv:astro-ph/0210603];
  V.~Acquaviva, N.~Bartolo, S.~Matarrese and A.~Riotto,
  Nucl.\ Phys.\  B {\bf 667}, 119 (2003)
  [arXiv:astro-ph/0209156].

\bibitem{Dutta:2007cr}
  B.~Dutta, J.~Kumar and L.~Leblond,
  JHEP {\bf 0707}, 045 (2007)
  [arXiv:hep-th/0703278].

\bibitem{Binetruy:1996xj}
  P.~Binetruy and G.~R.~Dvali,
  Phys.\ Lett.\  B {\bf 388}, 241 (1996)
  [arXiv:hep-ph/9606342];
  E.~Halyo,
  Phys.\ Lett.\  B {\bf 387} (1996) 43
  [arXiv:hep-ph/9606423].

\bibitem{Bernardeau:2007xi}
  F.~Bernardeau and T.~Brunier,
  Phys.\ Rev.\  D {\bf 76}, 043526 (2007)
  [arXiv:0705.2501 [hep-ph]].

\bibitem{Bartolo:2001cw}
  N.~Bartolo, S.~Matarrese and A.~Riotto,
  Phys.\ Rev.\  D {\bf 65}, 103505 (2002)
  [arXiv:hep-ph/0112261];
  F.~Bernardeau and J.~P.~Uzan,
  Phys.\ Rev.\  D {\bf 66}, 103506 (2002)
  [arXiv:hep-ph/0207295];
  F.~Bernardeau and J.~P.~Uzan,
  Phys.\ Rev.\  D {\bf 67}, 121301 (2003)
  [arXiv:astro-ph/0209330].

\bibitem{Sasaki:2008uc}
  L.~Alabidi and D.~Lyth,
  JCAP {\bf 0608}, 006 (2006)
  [arXiv:astro-ph/0604569];
  L.~Alabidi,
  JCAP {\bf 0610}, 015 (2006)
  [arXiv:astro-ph/0604611];
  M.~Sasaki,
  arXiv:0805.0974 [astro-ph].

\bibitem{Dvali:2003em}
  G.~Dvali, A.~Gruzinov and M.~Zaldarriaga,
  Phys.\ Rev.\  D {\bf 69}, 023505 (2004)
  [arXiv:astro-ph/0303591];
  G.~Dvali, A.~Gruzinov and M.~Zaldarriaga,
  Phys.\ Rev.\  D {\bf 69}, 083505 (2004)
  [arXiv:astro-ph/0305548].

\bibitem{Bevis:2007gh}
  R.~A.~Battye, B.~Garbrecht and A.~Moss,
  JCAP {\bf 0609}, 007 (2006)
  [arXiv:astro-ph/0607339].
  N.~Bevis, M.~Hindmarsh, M.~Kunz and J.~Urrestilla,
  Phys.\ Rev.\ Lett.\  {\bf 100}, 021301 (2008)
  [arXiv:astro-ph/0702223].

\bibitem{Sasaki:1995aw}
  M.~Sasaki and E.~D.~Stewart,
  Prog.\ Theor.\ Phys.\  {\bf 95}, 71 (1996)
  [arXiv:astro-ph/9507001].

\bibitem{Lyth:2001nq}
  A.~D.~Linde and V.~F.~Mukhanov,
  Phys.\ Rev.\  D {\bf 56}, 535 (1997)
  [arXiv:astro-ph/9610219];
  D.~H.~Lyth and D.~Wands,
  Phys.\ Lett.\  B {\bf 524}, 5 (2002)
  [arXiv:hep-ph/0110002].

\bibitem{Buchbinder:2007at}
  E.~I.~Buchbinder, J.~Khoury and B.~A.~Ovrut,
  arXiv:0710.5172 [hep-th];
  J.~L.~Lehners and P.~J.~Steinhardt,
  Phys.\ Rev.\  D {\bf 77}, 063533 (2008)
  [arXiv:0712.3779 [hep-th]];
  J.~L.~Lehners and P.~J.~Steinhardt,
  arXiv:0804.1293 [hep-th].


\bibitem{Lyth:2004gb}
  D.~H.~Lyth, K.~A.~Malik and M.~Sasaki,
  JCAP {\bf 0505}, 004 (2005)
  [arXiv:astro-ph/0411220];
  D.~H.~Lyth and Y.~Rodriguez,
  Phys.\ Rev.\ Lett.\  {\bf 95}, 121302 (2005)
  [arXiv:astro-ph/0504045].

\bibitem{falk}
T.~Falk, R.~Rangarajan and M.~Srednicki,
 Astrophys.\ J.\  {\bf 403} (1993) L1
 [arXiv:astro-ph/9208001].

\bibitem{again}
 M.~Zaldarriaga,
 Phys.\ Rev.\  D {\bf 69} (2004) 043508
 [arXiv:astro-ph/0306006];
  D.~Seery, K.~A.~Malik and D.~H.~Lyth,
 JCAP {\bf 0803} (2008) 014
[arXiv:0802.0588 [astro-ph]].

\bibitem{Babich:2004gb}
  D.~Babich, P.~Creminelli and M.~Zaldarriaga,
  JCAP {\bf 0408}, 009 (2004)
  [arXiv:astro-ph/0405356].

\bibitem{Chen:2006nt}
  D.~Seery and J.~E.~Lidsey,
  JCAP {\bf 0506} (2005) 003
  [arXiv:astro-ph/0503692].
 X.~Chen, M.~x.~Huang, S.~Kachru and G.~Shiu,
  JCAP {\bf 0701}, 002 (2007)
  [arXiv:hep-th/0605045].

\bibitem{Barnaby:2006hi}
  N.~Barnaby, T.~Biswas and J.~M.~Cline,
  JHEP {\bf 0704}, 056 (2007)
  [arXiv:hep-th/0612230].
  N.~Barnaby and J.~M.~Cline,
  JCAP {\bf 0707}, 017 (2007)
  [arXiv:0704.3426 [hep-th]].
  N.~Barnaby and J.~M.~Cline,
  arXiv:0802.3218 [hep-th].

\bibitem{Lyth:2007jh}
  D.~H.~Lyth,
  JCAP {\bf 0712}, 016 (2007)
  [arXiv:0707.0361 [astro-ph]].

\bibitem{Crocce:2005xy}
  M.~Crocce and R.~Scoccimarro,
  Phys.\ Rev.\  D {\bf 73}, 063519 (2006)
  [arXiv:astro-ph/0509418].
  C.~T.~Byrnes, K.~Koyama, M.~Sasaki and D.~Wands,
  JCAP {\bf 0711}, 027 (2007)
  [arXiv:0705.4096 [hep-th]].
  D.~Seery,
  JCAP {\bf 0802}, 006 (2008)
  [arXiv:0707.3378 [astro-ph]].

\bibitem{Zaballa:2006pv}
  I.~Zaballa, Y.~Rodriguez and D.~H.~Lyth,
  JCAP {\bf 0606}, 013 (2006)
  [arXiv:astro-ph/0603534].

\bibitem{Leblond:2008gg}
  L.~Leblond and S.~Shandera,
  arXiv:0802.2290 [hep-th].

\bibitem{Pogosian:2003mz}
  L.~Pogosian, S.~H.~H.~Tye, I.~Wasserman and M.~Wyman,
  Phys.\ Rev.\  D {\bf 68}, 023506 (2003)
  [Erratum-ibid.\  D {\bf 73}, 089904 (2006)]
  [arXiv:hep-th/0304188].
  
\bibitem{Enqvist:2005qu}
  K.~Enqvist, A.~Jokinen, A.~Mazumdar, T.~Multamaki and A.~Vaihkonen,
  JCAP {\bf 0503}, 010 (2005)
  [arXiv:hep-ph/0501076].

\bibitem{Leblond:2006cc}
  D.~H.~Lyth and A.~Riotto,
  Phys.\ Rev.\ Lett.\  {\bf 97}, 121301 (2006)
  [arXiv:astro-ph/0607326];
  L.~Leblond and S.~Shandera,
  JCAP {\bf 0701}, 009 (2007)
  [arXiv:hep-th/0610321].

\bibitem{Haack:2008yb}
  M.~Haack, R.~Kallosh, A.~Krause, A.~Linde, D.~Lust and M.~Zagermann,
  arXiv:0804.3961 [hep-th].

\bibitem{Brandenberger:2008if}
  R.~H.~Brandenberger, K.~Dasgupta and A.~C.~Davis,
  arXiv:0801.3674 [hep-th].

\end{thebibliography}
\end{document}